\documentclass[bjps]{imsart}

\RequirePackage{amsthm,amsmath,amsfonts,amssymb}
\RequirePackage[authoryear]{natbib}
\RequirePackage[colorlinks,citecolor=blue,urlcolor=blue]{hyperref}
\RequirePackage{graphicx}

\usepackage{verbatim}
\usepackage{amsmath}
\usepackage{graphicx,psfrag,epsf}
\usepackage{enumerate}
\usepackage{natbib}
\usepackage{multirow}
\usepackage{epstopdf}
\usepackage{array}
\usepackage{url} 
\usepackage{scalefnt} 
\usepackage[shortlabels]{enumitem} 

\startlocaldefs
\theoremstyle{plain}

\theoremstyle{remark}

\newcommand{\mat}[1]{\mbox{\boldmath{$#1$}}}

\numberwithin{equation}{section}

\endlocaldefs

\begin{document}






\begin{frontmatter}
\title{On variable selection in joint modeling of mean and dispersion}

\runtitle{Variable selection in joint modeling of mean and dispersion}

\begin{aug}
\author[A]{\inits{E. R.}\fnms{Edmilson Rodrigues} \snm{Pinto}\ead[label=e1]{edmilson.pinto@ufu.br}}\thanksref{t1}
\and
\author[A]{\inits{L. A.}\fnms{Leandro Alves} \snm{Pereira}\ead[label=e2,mark]{leandro.ap@ufu.br}}\thanksref{t1}

\thankstext{t1}{The authors thank Fapemig for financial support.}

\runauthor{Pinto, E.R. and Pereira, L.A.}

\address[A]{Federal University of Uberlândia, Uberlândia, Brazil,
\printead{e1,e2}}
\end{aug}

\begin{abstract}
The joint modeling of mean and dispersion (JMMD) provides an efficient method to obtain useful models for the mean and dispersion, especially in problems of robust design experiments. However, in the literature on JMMD there are few works dedicated to variable selection and this theme is still a challenge. In this article, we propose a procedure for selecting variables in JMMD, based on hypothesis testing and the quality of the model's fit. A criterion for checking the goodness of fit is used, in each iteration of the selection process, as a filter for choosing the terms that will be evaluated by a hypothesis test. Three types of criteria were considered for checking the quality of the model fit in our variable selection procedure. The criteria used were: the extended Akaike information criterion, the corrected Akaike information criterion and a specific criterion for the JMMD, proposed by us, a type of extended adjusted coefficient of determination. Simulation studies were carried out to verify the efficiency of our variable selection procedure. In all situations considered, the proposed procedure proved to be effective and quite satisfactory. The variable selection process was applied to a real example from an industrial experiment. 

\end{abstract}

\begin{keyword}
\kwd{Model selection}
\kwd{location-dispersion analysis}
\kwd{coefficient of determination}
\kwd{robust design experiment}
\kwd{generalized linear models}
\kwd{extended quasi-likelihood}
\end{keyword}

\end{frontmatter}

\section{Introduction} 
\label{sec:intro}

The variable selection process that will be considered in this article concerns the joint modeling of the mean and dispersion (JMMD), which is an extension of the generalized linear models (GLMs), introduced by \cite{NelderWedderburn}.

GLMs assume an exponential family distribution for the response variable and are more general than normal linear methods in that a mean-variance relationship appropriate for the data can be accommodated and an appropriate scale can be chosen for modeling the mean on which the action of the covariates is approximately linear. In generalized linear models the focus is on modeling and estimating the mean structure of the data while treating the dispersion parameter as a constant since GLMs automatically allow for dependence of the variance on the mean through the distributional assumptions. However, there are situations in which the observed data may exhibit greater variability than the one which is implied by the mean-variance relationship and thus the loss of efficiency in estimating the mean parameters, using constant dispersion models when the dispersion is varying, may be substantial \citep{AntoniadisEtal}.  

The joint modeling of the mean and dispersion is a general method that overcomes this difficulty, allowing the modeling of dispersion as a function of covariates, by the use of two interlinked generalized linear models. JMMD has enormous potential for application in science and industry and therefore has attracted the attention of many researchers. In the following, we present some relevant works in this area. \cite{Aitkin} considered a log-linear regression model for the variance in Gaussian models; \cite{NelderPregibon} introduced the extended quasi-likelihood (EQL); \cite{DavidianCarroll1} used the pseudo-likelihood method (PL), see \cite{CarrollRuppert}, p. 71; \cite{DavidianCarroll2} showed the asymptotic equivalence between EQL and PL and that EQL can be affected by an asymptotic bias when the underlying distribution is asymmetric and outside the exponential family. \cite{Smyth} proposed the class of double generalized linear models (DGLMs) for modeling of mean and dispersion; further developments for EQL were provided by \cite{GodambeThompson} and \cite{McCullaghNelder}; \cite{NelderLee} introduced the JMMD, using EQL, for robust parameter design, see \cite{Taguchi}; \cite{NelderLee1} compared EQL and PL in simulation studies with finite samples and showed that the maximum extended quasi-likelihood estimator is usually superior than the pseudo-likelihood estimator in minimizing the mean-squared error, see also \cite{Nelder}; \cite{Verbyla} used restricted maximum likelihood (REML) for  normal heteroscedastic models; \cite{LeeNelder0} introduced the class of hierarchical generalized linear models (HGLMs) as a synthesis of JMMD, generalizing the ideas of quasi-likelihood and EQL, see \cite{LeeNelderIII} and \cite{LeeNelderPawitan}; \cite{LeeNelderI} introduced the adjustment for the REML in the JMMD proposed by \cite{NelderLee}; \cite{SmythVerbyla1} and \cite{SmythVerbyla2} considered extensions of REML estimation of variance parameters for DGLMs; \cite{LeeNelderII} showed that the double exponential family, see \cite{Efron}, and the EQL generate identical inferences; \cite{CuervoGamerman} presented a Bayesian version of JMMD for normal heteroscedastic linear models; \cite{AntoniadisEtal} considered estimation and variable selection for JMMD in proper dispersion models, see \cite{Jorgensen0} and \cite{Jorgensen2}; \cite{BonatJorgensen} proposed the class of multivariate covariance generalized linear models (MCGLMs). 

In this article, the approach considered for the JMMD uses the adjusted EQL, as proposed by \cite{LeeNelderI}. Although the JMMD can be applied in several areas, one of its main applications and, therefore, the focus of this article, is in robust design experiments (RDE), a very useful tool for improving quality in industrial experiments, see \cite{Phadke} and \cite{Nair_Ed}. According to \cite{NelderLee1}, for data sets in which the dispersion must be modeled with reasonably small amounts of data, as in RDE, the best optimization criterion for model fitting in the JMMD is the EQL. 

In RDE, modeling and controlling the variance is of paramount importance, however this task is hampered by the fact that the experimental data are, in general, obtained from unreplicated fractional factorial experiments with few experimental runs. The JMMD allows the dispersion to be modeled very effectively in RDE problems, even without replication and the variance model can be obtained from the dispersion model in a very simple way. Given the importance of modeling the mean and the variability in industrial processes, if we are considering the JMMD, the question that arises is how to find the best model for mean and the best model for dispersion.

The methods used to select variables in the JMMD applied to RDE are not very clear and are based mainly on graphical analysis, see, for example, \cite{EngelHuele}. On the other hand, to carry out a formal process of selection of variables, we would have to consider a large number of possible models, which makes the process quite laborious. Thus, a clear, reliable and efficient method of selecting variables is of fundamental importance.

The aim of this article is to propose a consistent methodology that allows the selection of variables for both the mean and dispersion models in the JMMD. We propose a procedure for selecting variables, based on hypothesis testing and the quality of the model’s fit.
A criterion for checking the goodness of fit is used, in each iteration of the selection process, as a filter for choosing the terms that will be evaluated by a hypothesis test. Three types of criteria were considered for checking the quality of the model fit in our variable selection procedure. The criteria used were: the extended Akaike information criterion, the corrected Akaike information criterion and a specific criterion for the JMMD, proposed by us, a type of extended adjusted coefficient of determination.

The selection of variables for joint models of the mean and dispersion was considered by \cite{WangZhang}, where was proposed an extension of the Akaike information criterion based on EQL; \cite{WuLi}, for the normal inverse distribution; \cite{WuZhangXu}, for lognormal distribution; \cite{CharalambousPanTranmer}, for multilevel models; \cite{CharalambousPanTranmer1}, for HGLMs;  \cite{AntoniadisEtal}, for proper dispersion models; \cite{BayerCribariNeto}, for beta regression models; \cite{WuLiTao}, for mixture of regression models.

The article is organized as follows. In Section \ref{JMMD}, the joint modeling of mean and dispersion is briefly described. In Section \ref{Criteria}, the proposed procedure and criterion are presented. In Section \ref{NumEvaluation}, simulation studies are carried out to evaluate the proposed procedure and criterion. An application to real data from an industrial experiment is presented in Section  \ref{Application}.  Finally, a conclusion is given in Section \ref{Conclusion}.

\section{Joint modeling of mean and dispersion}
\label{JMMD}

According to \cite{NelderLee}, the method of joint modeling of mean and dispersion consists of finding joint models for the mean and dispersion. In their approach, using the extended quasi-likelihood, two interlinked generalized linear models are needed, one for the mean and the other for the dispersion. 

Let $Y_1, \ldots,Y_n$ be $n$ independent random variables with the same probability distribution, representing the dependent response variables, whose observed values are given by $y_1,\ldots,y_n$.  For the $i$th response $Y_i$ it is assumed to be known that $E(Y_i)=\mu_i$ and $Var(Y_i)=\phi_iV(\mu_i)$, where $\phi_i$ is the dispersion parameter and $V(.)$ is the variance function in GLMs. The mean and dispersion models are built as follows.

Suppose $\mat{x}^t=(x_1,\ldots,x_s)$ and $\mat{z}^t=(z_1,\ldots,z_r)$ are the vectors of the independent variables that affect the mean and dispersion models, respectively. The vectors $\mat{x}$ and $\mat{z}$ may contain, in addition to design factors that are controlled by the experimenter, noise factors that cannot be controlled and are considered as random variables. 

Let $\varphi$ be a link function for the mean model, i.e., for the $i$th response, the linear predictor for the mean model is given by $\eta_i=\varphi(\mu_i)=\mat{f}^t(\mat{x}_i)\mat{\beta}$ with $\mat{f}^t(\mat{x}_i)=(f_1(\mat{x}_i),\ldots,f_p(\mat{x}_i))$ where $f_j(\mat{x}_i)$, for $j=1,\ldots,p$, is a known function of $\mat{x}_i$ and $\mat{\beta}$ is a $p\times1$ vector of unknown parameters.

For the dispersion model is used as response variable the deviance component, which, for each observation $y_i$, is given by

\begin {equation} \label{eq3}
d_{i} = 2 \int_{\mu_{i}}^{y_{i}}\frac{(y_i- t)}{V(t)}dt,   
\end{equation}

\noindent see \cite{McCullaghNelder}, p. 360.

Following \cite{LeeNelderI}, for the dispersion model we are assuming a Gamma model with a log link function, i.e., for the $i$th response, the linear predictor for the dispersion model is given by $\xi_i=\log(\phi_i)=\mat{g}^t(\mat{z}_i)\mat{\gamma}$, with $\mat{g}^t(\mat{z}_i)=(g_1(\mat{z}_i),\ldots,g_q(\mat{z}_i))$, where $g_j(\mat{z}_i)$, for $j=1,\ldots,q$, is a known function of $\mat{z}_i$ and $\mat{\gamma}$ is a $q\times1$ vector of unknown parameters. We also define $\mat{X}=[\mat{f}(\mat{x}_1),\ldots,\mat{f}(\mat{x}_n)]^t$ the $n \times p$ experimental matrix for the mean model and $\mat{Z}=[\mat{g}(\mat{z}_1),\ldots,\mat{g}(\mat{z}_n)]^t$ the $n \times q$ experimental matrix for the dispersion model.

The fitting for the JMMD uses as an optimizing criterion the extended quasi-likelihood, introduced by \cite{NelderPregibon}, see \cite{McCullaghNelder}, p. 349. In this work we use the adjusted extended quasi-likelihood, introduced by \cite{LeeNelderI} and given by

\begin{equation} \label{eq4}
Q^{+}(\mat{\mu},\mat{\phi};\mat{y})=\sum_{i=1}^{n} -\frac{1}{2}\left( \frac{d_{i}^{*}}{\phi_{i}} + \log \{ 2\pi\phi_{i}V(y_{i})\}\right)
\end{equation}

\noindent where $d_{i}^{*}=\frac{d_{i}}{1-h_{i}}$ is the standardized deviance component and $h_{i}$ is the $i$th element of the diagonal of the matrix $\mat{H}=\mat{W}^{\frac{1}{2}}\mat{X}(\mat{X}^{t}\mat{WX})^{-1}\mat{X}^{t}\mat{W}^{\frac{1}{2}}$, being $\mat{W}$, the $n \times n$ weight matrix for the GLMs, a diagonal matrix with elements given by $w_{i}=\left(\frac{\partial\mu_{i}}{\partial\eta_{i}}\right)^{2}\frac{1}{\phi_i V(\mu_{i})}$ with $i=1,\ldots,n$. The Table \ref{Tab1} gives a resume of the joint modeling of mean and dispersion. From Table \ref{Tab1}, we can observe that the standardized deviance component from the model for the mean becomes the response for the dispersion model, and the inverse of fitted values for the dispersion model provides the prior weights for the mean model.

\begin {table}[htb]
\caption{Summary of the JMMD for the {\it i}th response}
\begin{tabular}{p{3.5cm} p{3.5cm} p{3.5cm}}
\hline
Component               & Mean model                                           & Dispersion model$^{\dag}$                        \\ \hline
Response variable       &  $y_i$                                               &  $d_{i}^{*}$                                     \\
Mean                    &  $\mu_i$                                             &  $\phi_i$                                        \\ 
Variance                &  $\phi_i V(\mu_i)$                                   &  $2\phi_{i}^{2}$                                 \\
Link function           &  $\eta_i=\varphi(\mu_i)$                             &  $\xi_i=\log(\phi_i)$                             \\
Linear predictor        &  $\eta_i=\mat{f}^t(\mat{x}_i)\mat{\beta}$            &  $\xi_i=\mat{g}^t(\mat{z}_i)\mat{\gamma}$        \\
Deviance component      &  $d_i=2 \int_{\mu_i}^{y_i}\frac{y_i-t}{V(t)}dt$     &  
$2\left\{-\log\left(\frac{d_{i}^{*}}{\phi_i}\right) + \frac{(d_{i}^{*}-\phi_i)}{\phi_i}\right\}$     \\
Prior weight            &  $\frac{1}{\phi_i}$                                  &  $(1-h_i)/2$        \\ \hline
\multicolumn{3}{l}{$^{\dag}$\scriptsize{For the dispersion model we are assuming a Gamma model with logarithmic link function}}
\end{tabular}
\label{Tab1}
\end{table}

The algorithm for parameter estimation is an extension of the standard GLMs algorithm, in which the model for the mean is fitted assuming that the fitted values for the dispersion are known and that the model for dispersion is fitted using the fitted values for the mean. The fitting alternates between the mean and dispersion models until convergence is achieved.

In the next section, we present a procedure for variable selection in JMMD, considering two different criteria to verify the goodness of the model fit.

\section{Procedure for variable selection in JMMD}
\label{Criteria}

For our procedure of variable selection in the joint modeling of the mean and dispersion we consider two alternative criteria to verify the goodness of the model fit. The first criterion, proposed by us, is a combination of the criteria proposed by \cite{HuShao} and \cite{Zhang} for selection of variables in generalized linear models.
\cite{HuShao} propose a criterion of goodness of fit given by

\begin{equation} 
  \label{eqn_r2HuShao}
\footnotesize{  
  R^2_{HS}= 1 - \frac{\sum_{i=1}^{n} (y_i -\widehat{\mu}_i)^2/(n-\lambda_{n}r) }{\sum_{i=1}^{n} (y_i -\overline{y})^2/(n-1)}},
\end{equation}

\noindent where $\overline{y}=\frac{1}{n}\sum_{i=1}^{n}y_i$, $\widehat{\mu}_i=\varphi^{-1}(\widehat{\eta}_i)$, $n$ is the number of observations, $r$ is the number of estimated parameters of the model and $\lambda_n$ is some function of $n$, satisfying $\lambda_n \rightarrow \infty$ and $\lambda_n/n \rightarrow 0$ as the sample size $n \rightarrow \infty$. The $\lambda_n$ functions considered by them in their simulation studies were $\sqrt{n}$ and $\log n$. 

The adjusted criterion proposed by \cite{Zhang} to verify the goodness of the model fit is given by

\begin{equation} 
  \label{eqn:r2Zhang} 
\footnotesize{  
  R^2_{Z}= 1- \frac{\sum_{i=1}^{n}  d_{V}(y_i,\widehat{\mu}_i)/(n-r) }{\sum_{i=1}^{n} d_{V}(y_i,\overline{y})/(n-1)}},
\end{equation} 

\noindent where $d_{V}(a,b)= \left[ \int_{a}^{b}\sqrt{1+[V'(t)]^2}dt \right]^2$ with $V'(t)=\frac{dV(t)}{dt}$ and $V(t)$ is a continuous and derivable function in $(a, b)$. The function $d_V(a, b)$ is the squared arc length of the variance function between $V(a)$ and $V(b)$. In equation (\ref{eqn:r2Zhang}), $\sum_{i=1}^{n}  d_{V}(y_i,\widehat{\mu}_i)$ represents the variation in the response unexplained by the model and $\sum_{i=1}^{n} d_{V}(y_i,\overline{y})$ represents the total variation in the response. 

According to \cite{Zhang}, $d_V(a,b)$ can differ dramatically from the Euclidean distance $(a-b)^2$ when the underlying variance function is nonlinear. For example, in a particular but rather general case of quadratic variance function, i.e., $V(\mu)=c_2\mu^2 + c_1\mu + c_0$, where $c_0$, $c_1$ and $c_2$ are known constants and $c_2 \neq 0$, $d_{V}(a,b)= \frac{1}{16c_2^2} \left\{ \log \left( \frac{\nu(b)} {\nu(a)} \right) + \nu(b) - \nu(a) \right\}^2$, where $\nu(t)=V'(t) + \sqrt{1+[V'(t)]^2}$. If $c_2 = 0$, the variance function is linear and $d_V (a, b)= (1+c_1^2)(a-b)^2$. Hence, for the Normal model, where $V(\mu) = 1$, $d_V(a,b)=(a-b)^2$; for the Poisson model, where $V(\mu)=\mu$, $d_V(a,b)=2(a-b)^2$.

The criterion proposed by \cite{Zhang} is well defined, as long as the underlying model, like quasi-models, specifies the mean and variance functions. In this way, $R_{Z}^{2}$ is applicable to quasi-likelihood models, because it does not depend on the complete specification of the likelihood function, requiring only to know the mean and the variance function.

In this way, the extension of the criterion proposed by \cite{Zhang} for joint models of the mean and dispersion is immediate, however, in the mean model, the dispersion parameter must be taken into account for the calculation of the criterion. In fact, as in the iterative process for the JMMD the variance for the {\it i}th observation is given by $Var(Y_i) = \phi_i V(\mu_i)$, where $\phi_i$ is known, given from the previous iteration. We can think of $\phi_i$ as a measure weight for $V(\mu_i)$.

Therefore, using the idea considered by \cite{Zhang} to measure the change of variation of the response variable, associated with the penalty function proposed by \cite{HuShao} and considering the dispersion parameter as weight for the variance function, we propose, for the process of joint modeling of the mean and dispersion, the following  criteria of goodness of fit. For the mean model we propose the criterion given by

\begin{equation} 
  \label{eqn:r2m} 
\footnotesize{  
  \tilde{R}_m^2= 1- \frac{\sum_{i=1}^{n} \tilde{d}_{V}(y_i,\widehat{\mu}_i)/(n-\lambda_{n}p) }{\sum_{i=1}^{n} \tilde{d}_{V}(y_i,\overline{y})/(n-1)}}
\end{equation}

\noindent and for the dispersion model the proposed criterion is

\begin{equation} 
  \label{eqn:r2d} 
\footnotesize{  
  \tilde{R}_d^2= 1- \frac{\sum_{i=1}^{n} d_{V}(d_i^{*},\widehat{\phi}_i)/(n-\lambda_{n}q) }{\sum_{i=1}^{n} d_{V}(d_i^{*},\overline{d^{*}})/(n-1)}},
\end{equation}

\noindent where $\tilde{d}_{V}(a,b)=\left[ \int_{a}^{b}\sqrt{1+ \phi^{2}[V'(t)]^2}dt \right]^2$ is calculated for the mean model as proposed by \cite{Zhang}, but considering the dispersion parameter as weight for the variance function; $d_V(a,b)$, calculated for the dispersion model, is the same as that given in equation (\ref{eqn:r2Zhang}); $n$ is the number of observations, $\lambda_n$ is a function of $n$, as proposed by \cite{HuShao}, i.e., $\lambda_n \rightarrow \infty$ and $\lambda_n/n \rightarrow 0$ as the sample size $n \rightarrow \infty$, $p$ is the number of parameters in the mean model, $q$ is the number of parameters in the dispersion model, $\hat{\mu}_i=\varphi^{-1}(\hat{\eta}_i)$, $\hat{\phi}_i=\exp(\hat{\xi}_i)$, $\overline{y}=(1/n)\sum_{i=1}^n y_{i}$ and $\overline{d^{*}}=(1/n)\sum_{i=1}^n d_{i}^{*}$, with $y_i$ and $d_i^{*}$ given in Table \ref{Tab1}. In equation (\ref{eqn:r2d}) we can notice that if $\lambda_n = 1$ we have the criterion proposed by \cite{Zhang}. 

Note that, as in the criterion proposed by \cite{Zhang}, $\sum_{i=1}^{n}  \tilde{d}_{V}(y_i,\widehat{\mu}_i)$ and $\sum_{i=1}^{n}  d_{V}(d_i^*,\widehat{\phi}_i)$  represent the variation in the response unexplained by the model and $\sum_{i=1}^{n} \tilde{d}_{V}(y_i,\overline{y})$  and $\sum_{i=1}^{n} d_{V}(d_i^*,\overline{d^*})$ represent the total variation in the response. Thus, for both $\tilde{R}_m^2$ and $\tilde{R}_d^2$ criteria, the higher the value, the better. It is also worth mentioning that, depending on the models considered, including their number of parameters, the sample size and the choice of the $\lambda_n$  function, the $\tilde{R}_m^2$ and $\tilde{R}_d^2$ criteria may have values outside the range $[0,1]$.

In equation (\ref{eqn:r2d}), since we are considering the Gamma model for the dispersion, $d_{V}(a,b)= \frac{1}{16} \left\{ \log \left( \frac{2b + \sqrt{1+4b^2}} {2a + \sqrt{1+ 4a^2}} \right) + 2b\sqrt{1 + 4b^2} - 2a\sqrt{1 + 4a^2} \right\}^2$. For the mean models that will be considered in this article the values of $\tilde{d}_{V}(a,b)$, used in equation (\ref{eqn:r2m}), are calculated by:  $\tilde{d}_{V}(a,b)=(b-a)^2$ for the Normal type model; $\tilde{d}_{V}(a,b)=(1+\phi^2)(b-a)^2$ for the Poisson type model and for the Binomial type model, $\tilde{d}_{V}(a,b)= \frac{1}{16\phi^2} \left\{ \log \left( \frac{\beta + \sqrt{1+\beta^2}} {\alpha + \sqrt{1+\alpha^2}} \right) + \beta\sqrt{1+\beta^2} - \alpha\sqrt{1+\alpha^2} \right\}^2$, where $\alpha=\phi(1-2a)$ and $\beta=\phi(1-2b)$. Note that in the case of the Normal type model our criterion is the same criterion proposed by \cite{HuShao}.

The second criterion, alternative to the first, considered for the mean model in our procedure for selection of variables in JMMD, is the extended Akaike information criterion ($EAIC$), proposed by \cite{WangZhang} and given by

\begin{equation}
\label{eq.eaic}
 EAIC=-2Q^{+}(\mat{\mu},\mat{\phi};\mat{y}) + F(\kappa,n),
\end{equation}

\noindent where $\kappa=p+q$ is the sum of the number of parameters of the mean and dispersion models and $F(\kappa,n)$ is a  penalty function that depends on $\kappa$ and $n$. In our simulation studies we consider $F(\kappa,n)=\frac{2 \kappa n}{n-\kappa-1}$. Such penalty function, originally proposed by \cite{Sugiura} as a bias  correction  to the  Akaike information criterion in the context of linear regression models, has been extended to nonlinear regression and autoregressive time series models by  \cite{HurvichTsai}. For the dispersion model, as we are considering the known likelihood, the criterion used will be the corrected Akaike information criterion ($AIC_c$), given by $AIC_c = -2\ln \hat{L} + \frac{2qn}{n-q-1}$, where $\hat{L}$ is the maximum value of the likelihood function for the Gamma dispersion model. For both $EAIC$ and $AIC_c$ criteria, the lower the value, the better. 

The Algorithm 1 shows how the terms of the model considered (mean or dispersion) are selected in order to find the best model. \vspace{0.2cm}

\noindent {\bf Algorithm 1}: Pseudo algorithm for variable selection in mean (dispersion) model \vspace{0.2cm}

In the JMMD, for a given and fixed dispersion (mean) model, choose the terms of the mean (dispersion) model, considering the following sequence of steps.

   \begin{enumerate}
   \item Let $\cal{V}$ be the set of terms that will be used in the variable selection process and consider $\vartheta = \theta_0$ the linear predictor of the initial model.
    \item Fit models with linear predictors $\vartheta_j=\vartheta +\theta_j v_j$, for all $v_j \in \cal{V}$. 
    \item For each model fitted in step 2, calculate the appropriate selection measure ($\tilde{R}_m^{2}$, $\tilde{R}_d^{2}$, $EAIC$, or $AIC_c$) to check the quality of fit and choose the model with linear predictor, say $\vartheta_k=\vartheta + \theta_k v_k$, that has the best fit. Evaluate the value of the selection measure found in the current iteration with the value of the selection measure found in the previous iteration. If the value of the current selection measure is better than the previous value, go to step 4, otherwise, go to step 5.
    
  \item For models with nested linear predictors $\vartheta$ and $\vartheta_k$, apply an appropriate test to verify the significance of the addition of $v_k$ into the linear predictor $\vartheta$. If $v_k$ is significant, remove the term $v_k$ from $\cal{V}$, do $\vartheta=\vartheta_k$ and return to step 2. Otherwise, the procedure ends and $\vartheta$ is the final model chosen.
  
   \item Evaluate, in the same way as in step 4, the significance of the addition of $v_k$ into the linear predictor $\vartheta$. If $v_k$ is significant, do $\vartheta=\vartheta_k$. Stop the procedure and $\vartheta$ is the final model chosen. 
   
  \end{enumerate} 
 
In step 4 of Algorithm 1, in the case of the dispersion model, for which we are assuming a GLM with Gamma distribution, the test for comparing two nested models is the usual analysis of the deviance for GLMs. In this case, if $H_a$ and $H_b$ are two nested hypothesis of dimension $a < b$, that is $\xi_a \subset \xi_b$, then, under $H_a$, the change in the deviance, given by $\frac{1}{2}\{D(\mat{d}^{*},\mat{\phi}_a) - D(\mat{d}^{*},\mat{\phi}_b)\}=\frac{1}{2}\left\{\sum_{i=1}^{n} d_{d_i}^{a} - \sum_{i=1}^{n} d_{d_i}^{b} \right\}$, has an asymptotic $\chi^2_{b-a}$, where
$D(\mat{d}^{*},\mat{\phi}_k)\}$ is the deviance for the model with mean $\phi_k$ and $d_{d_i}^{k}=2\left\{-\log\left(\frac{d_{i}^{*}}{\phi_{k_i}}\right) + \frac{(d_{i}^{*}-\phi_{k_i})}{\phi_{k_i}}\right\}$ is the deviance component, as shown in Table 1. In the case of the mean model, where $\mat{\phi}$ is given, the test for comparing two nested models is obtained by the quasi-likelihood ratio test applied to the extended quasi-likelihood. The justification for using the quasi-likelihood ratio test in the mean model is given by the fact that the extended quasi-likelihood is a saddle point approximation for the log likelihood, especially when there  exists a distribution  of  the exponential  family  with  a given  variance  function, see \cite{NelderLee} and \cite{McCullaghNelder}, p. 350. Thus, in the same way as we considered previously in the case of the dispersion model, if $H_c$ and $H_d$ are two nested hypothesis of dimension $c < d$, that is $\eta_c \subset \eta_d$, then, under $H_c$, the change in the extended quasi-deviance, given by $-2\left\{ Q^{+}(\mat{\mu}_c,\mat{\phi};\mat{y}) - Q^{+}(\mat{\mu}_d,\mat{\phi};\mat{y})\right\}= \sum_{i=1}^{n}\frac{1}{\phi_i} d^{*}_{\mu_{c_i}} - \sum_{i=1}^{n}\frac{1}{\phi_i} d^{*}_{\mu_{d_i}}$, has an asymptotic $\chi^2_{d-c}$, where $d^{*}_{\mu_{k_i}}$ is the standardized deviance component in relation to $\mu_{k_i}$, given in Table 1. However, this result is true when $\phi_i$ is the correct value, which is not the case in our iterative learning process for constructing joint mean and dispersion models. Therefore, for our procedure, due to uncertainty in the estimation of $\phi_i$, we assume that $\phi_i$ is replaced by $\tau \phi_i$, where $\tau$ is an unknown constant. It can be said that $\tau$ is a compensation for not being able to estimate $\phi_i$ exactly. Consequently, the hypothesis test considered in our mean model selection procedure, which overcomes the dependency on the unknown parameter $\tau$, is the {\it F}-test and it is not difficult to verify that 

\begin{equation}
\label{eqn:F}  
\frac{\left(\sum_{i=1}^{n}\frac{1}{\phi_i} d^{*}_{\mu_{c_i}} - \sum_{i=1}^{n}\frac{1}{\phi_i} d^{*}_{\mu_{d_i}} \right) \Big/ (d-c)}{\left(\sum_{i=1}^{n}\frac{1}{\phi_i} d^{*}_{\mu_{d_i}} \right) \Big/(n-d)}
\end{equation}

\noindent has an asymptotic $F_{d-c,n-d}$. For a comprehensive understanding of the relationship between the {\it F}-test and the likelihood ratio test, considering the saddle point approximation, see \cite{Jorgensen1}, p. 89.

Our selection strategy, shown in Algorithm 2, consists of a three-step scheme in which the selected dispersion model is used to select the best mean model and vice versa. In our selection scheme, we use a recursive procedure that is only finalized when the value used as a measure of goodness of fit for the mean model reaches an optimal value, for example, when $\tilde{R}_m^2$ stops increasing or when $EAIC$ stops decreasing. \vspace{0.2cm}

\noindent {\bf Algorithm 2}: Pseudo algorithm for variable selection in JMMD \vspace{0.2cm}

\begin{enumerate} 

\item Assuming constant dispersion, use Algorithm 1 to find the terms of the best current mean model and for this model calculate the value of $\tilde{R}_m^2$ ($EAIC$).

\item Assuming that the selected current mean model is adequate, use Algorithm 1 to find the terms of the best current dispersion model.

\item Assuming that the selected current dispersion model is adequate, use Algorithm 1 to find the terms of the best current mean model and for this model calculate a new value of $\tilde{R}_m^2$ ($EAIC$). If the updated value of $\tilde{R}_m^2$ decreases ($EAIC$ increases) or is equal to the value previously obtained, stop and the models previously found for mean and dispersion are chosen. Otherwise, return to step 2.

\end{enumerate}

The selection scheme proposed in this article consists of a method that, in addition to providing a clear and objective process for selecting variables, aims to reduce computational costs inherent in the great complexity involved in variable selection for joint mean and dispersion models. In fact, in our variable selection process, if we assume that $m$ regressors are considered for each model in Algorithm 1, we have $\sum_{j=0}^{k-1} \binom{m-j}{1} + 1$ $=\frac{k}{2} (2m-k+1)+1$ possibilities of joint models to be evaluated, where $k \leq m$ represents the number of iterations of the algorithm. In our process, governed by Algorithm 2, Algorithm 1 is used twice, except in the first iteration, where it is used three times. Therefore, if we consider $l$ iterations of Algorithm 2, we would have to check $(2l+1)(\frac{k}{2} (2m-k+1)+1)$ joint models. Without using our process, we would have to evaluate $(\sum_{j=0}^{m} \binom{m}{j})\times (\sum_{j=0}^{m} \binom{m}{j})$ $=2^{m} \times 2^m=2^{2m}$ different joint models. As an illustration of this result, consider $l = m = k = 10$. In our procedure, we should evaluate $21 \times 56=1176$ joint models. Without using our procedure, we would have to evaluate $2^{20}=1048576$ joint models.

\section{Numerical evaluation}
\label{NumEvaluation}

To evaluate the accuracy of the variable selection procedure, Monte Carlo simulations were performed in order to verify the percentage of hits. The main objective is to evaluate how the variable selection scheme behaves for different sample sizes and how the criteria considered affect the proposed scheme.

In the evaluation process of the selected models, three possible situations were considered: category 1 (Type 1 model): when the selected model does not contain all the terms of the simulated model; category 2 (Type 2 model): when the selected model contains all the terms of the simulated model plus other terms and correct model (Optimal): when the selected model contains exactly the terms of the simulated model. All simulations were performed using the statistical software R \citep{RTeam}. 

The simulations were performed for mean models in which the response variable was considered to follow the Normal, Binomial or Poisson distribution. For the dispersion model, the response variable was considered to have a Gamma distribution. The general linear predictors, chosen for the data generation process, were:

\begin{equation}
  \label{eqn:m1}  
  \eta=\beta_0 + \beta_1x_1+\beta_2x_2+\beta_3x_3 
\end{equation} and

\begin{equation}
 \label{eqn:d1}   
  \xi=\gamma_0 + \gamma_1z_1+\gamma_2z_2+\gamma_3z_3, 
\end{equation}

\noindent where (\ref{eqn:m1}) was the linear predictor used for the mean model and (\ref{eqn:d1}) was the linear predictor used for the dispersion model. For the simulations, $16$ different scenarios were considered, which originated from the combination of four possibilities for the goodness of fit criteria with four different sample sizes. The sample sizes used in the simulations were $25$, $50$, $100$ and $150$. The criteria considered were the extended Akaike information criterion, the corrected Akaike information criterion  and three variants of our proposed criterion, one using $\lambda_n=1$, another with $\lambda_n=\log n$ and another with $\lambda_n=\sqrt{n}$. The extended Akaike information criterion was used in the mean model and the corrected Akaike information criterion was used in the dispersion model. For each scenario considered 1000 Monte Carlo replications were performed. The values of $\mat{x}^{t}=(x_{1},x_{2},x_{3})$ and $\mat{z}^{t}=(z_{1},z_{2},z_{3})$ were randomly obtained from a Uniform distribution between $-1$ and $1$. 

Regarding the level of significance, considered for the hypothesis tests in step 4 of Algorithm 1, we understand that a level of 0.05 is very strict and could exclude important variables from the model. \cite{LeeKoval} examined the question of the level of significance in stepwise procedures for the case of logistic regression and suggested that the level of significance adopted should be between 0.15 and 0.20. Based on the discussions presented by \cite{LeeKoval} and in the references considered there, we decided to use an intermediate value between 0.05 and 0.15, that is, a significance level of $0.10$. In all simulated studies performed, $i = 1,\ldots, n$ refers to the $i$th response observed and $n$ is the sample size. The simulation studies are shown below.

\subsection{Simulation study for Normal distribution}

For the mean model with Normal probability distribution or Normal type, i.e., with $V(\mu_i)=1$, the data simulation process must ensure that $E(Y_i)=\mu_i$ and $Var(Y_i)=\phi_i$. Such results can be directly obtained by taking $Y_i\sim Normal(\mu_i,\phi_i)$ or considering $S_i \sim \phi_i \chi^2_1$, where $\chi^2_1$ represents the Chi-square distribution with 1 degree of freedom,  and taking $Y_i =\pm \sqrt{S_i}+\mu_i$. 

The values chosen for the parameter vector $(\beta_0,\beta_1,\beta_2,\beta_3,$ $\gamma_0,\gamma_1,\gamma_2,\gamma_3)$ used to perform simulations were $(4,15,13,0,0.3,0,3,0)$, where the zero value means that the parameter was not used in the simulation. For each Monte Carlo simulation, the responses were generated considering log link function for the dispersion model and identity link function for the mean model. 

Table \ref{table:Freq} presents the percentage of Optimal, Type 1 and Type 2 models, obtained from 1000 Monte Carlo simulations of the proposed procedure of variable selection, as previously established at the beginning of Section \ref{NumEvaluation}. 

Analyzing the percentages for the mean models, referring to Optimal model in Table \ref{table:Freq}, we can see that the simulations performed with 25 sample size already have high success rates, but with considerable indices of Type 1 models. When we increased the sample size, we can observe the rapid increase in percentages for the Optimal model and a sharp drop in percentages for the Type 1 model. For the dispersion models, there is a lower rate of Optimal model, but an acceptable rate is achieved for samples of size 100 and 150. With respect to the Type 2 models, we noticed that for both mean and dispersion models the variable selection scheme presented low percentages and continued with low percentages even increasing the size of the sample.


\renewcommand\arraystretch{2}
\setlength\tabcolsep{2.8pt}


\begin{table}[htb!]
\scalefont{0.80}
\caption{Percentages of model types found in 1000 Monte Carlo simulations of the proposed variable selection procedure for the Normal distribution}
\label{table:Freq}

\begin{tabular}{ll|llccccccccccccclccccl}
\hline
                            &                            &  &  & \multicolumn{4}{c}{Optimal}                                                                               & \multicolumn{1}{l}{} & \multicolumn{1}{l}{} & \multicolumn{1}{l}{} & \multicolumn{4}{c}{Type 2}                                                                            & \multicolumn{1}{l}{} & \multicolumn{1}{l}{} &  & \multicolumn{4}{c}{Type 1}                                                                                &  \\ \cline{4-9} \cline{11-16} \cline{18-23} 
                            &                            &  &  & \multicolumn{4}{c}{Sample size}                                                                           & \multicolumn{1}{l}{} & \multicolumn{1}{l}{} & \multicolumn{1}{l}{} & \multicolumn{4}{c}{Sample size}                                                                       & \multicolumn{1}{l}{} & \multicolumn{1}{l}{} &  & \multicolumn{4}{c}{Sample size}                                                                           &  \\ \cline{5-8} \cline{12-15} \cline{19-22}
Model                       & Criterion                  &  &  & 25                       & 50                       & 100                      & 150                      &                      &                      &                      & 25                      & 50                      & 100                     & 150                     &                      &                      &  & 25                       & 50                       & 100                      & 150                      &  \\ \hline
\multirow{4}{*}{Mean}       & $\tilde{R}_{m}^2(\log n)$  &  &  & 71.0                     & 83.5                     & 92.0                     & 95.1                     &                      &                      &                      & 4.9                     & 5.1                     & 5.5                     & 4.8                     &                      &                      &  & 24.1                     & 11.4                     & 2.5                      & 0.1                      &  \\
                            & $\tilde{R}_{m}^2(\sqrt n)$ &  &  & 76.6                     & 88.5                     & 92.0                     & 98.9                     &                      &                      &                      & 1.0                     & 0.2                     & 5.5                     & 0.1                     &                      &                      &  & 22.4                     & 11.3                     & 2.5                      & 1.0                      &  \\
                            & $\tilde{R}_{m}^2(1)$       &  &  & \multicolumn{1}{l}{69.2} & \multicolumn{1}{l}{81.9} & \multicolumn{1}{l}{90.4} & \multicolumn{1}{l}{90.4} & \multicolumn{1}{l}{} & \multicolumn{1}{l}{} & \multicolumn{1}{l}{} & \multicolumn{1}{l}{5.8} & \multicolumn{1}{l}{7.0} & \multicolumn{1}{l}{5.9} & \multicolumn{1}{l}{8.4} & \multicolumn{1}{l}{} & \multicolumn{1}{l}{} &  & \multicolumn{1}{l}{25.0} & \multicolumn{1}{l}{11.1} & {3.7}  &{1.2}  &  \\
                            & $EAIC$                     &  &  & 66.6                     & 88.5                     & 91.7                     & 94.7                     &                      &                      &                      & 6.7                     & 0.2                     & 6.3                     & 5.0                     &                      &                      &  & 26.7                     & 11.3                     & 2.0                      & 0.3                      &  \\ \hline
\multirow{4}{*}{Dispersion} & $\tilde{R}_{d}^2(\log n)$  &  &  & 30.9                     & 53.4                     & 62.8                     & 68.8                     &                      &                      &                      & 0.8                     & 2.1                     & 4.5                     & 4.0                     &                      &                      &  & 68.3                     & 44.5                     & 32.7                     & 27.2                     &  \\
                            & $\tilde{R}_{d}^2(\sqrt n)$ &  &  & 32.0                     & 53.8                     & 62.9                     & 67.5                     &                      &                      &                      & 1.2                     & 3.5                     & 4.4                     & 3.4                     &                      &                      &  & 66.8                     & 42.7                     & 32.7                     & 29.1                     &  \\
                            & $\tilde{R}_{m}^2(1)$       &  &  & \multicolumn{1}{l}{30.9} & \multicolumn{1}{l}{52.1} & \multicolumn{1}{l}{63.8} & \multicolumn{1}{l}{67.9} & \multicolumn{1}{l}{} & \multicolumn{1}{l}{} & \multicolumn{1}{l}{} & \multicolumn{1}{l}{1.8} & \multicolumn{1}{l}{3.6} & \multicolumn{1}{l}{2.7} & \multicolumn{1}{l}{1.5} & \multicolumn{1}{l}{} & \multicolumn{1}{l}{} &  & \multicolumn{1}{l}{67.3} & \multicolumn{1}{l}{44.3} & \multicolumn{1}{l}{33.5} & \multicolumn{1}{l}{30.6} &  \\
                            & $AIC_c$                      &  &  & 36.1                     & 58.1                     & 71.9                     & 78.2                     &                      &                      &                      & 1.6                     & 5.6                     & 7.0                     & 4.0                     &                      &                      &  & 62.3                     & 36.3                     & 21.1                     & 17.8                     &  \\ \hline
\end{tabular}
\end{table}

\begin{figure}[!ht]
\centering
\includegraphics[scale=0.7]{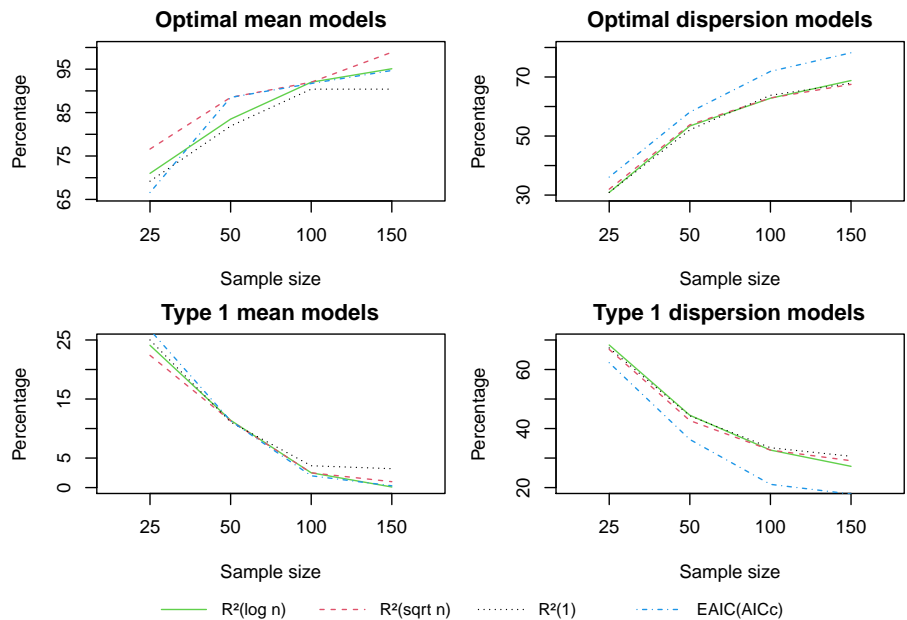}
\caption{Graphs, obtained from the simulations of the Normal distribution, regarding the relationship between sample size and percentage found for the Optimal and Type 1 models, for the $\tilde{R}_{m}^2$, $\tilde{R}_{d}^2$, $EAIC$ and $AIC_c$ criteria,  where the solid line represents the graph for $\tilde{R}_{m}^2$ $(\tilde{R}_{d}^2)$ with $\lambda_n=\log n$, the dashed line represents the graph for $\tilde{R}_{m}^2$ $(\tilde{R}_{d}^2)$, with $\lambda_n=\sqrt n$, the dotted line represents the graph for $\tilde{R}_{m}^2$ $(\tilde{R}_{d}^2)$, with $\lambda_n=1$ and the dash-dotted line represents the graph for $EAIC (AIC_c)$.}
\label{fig:graphics-normal}
\end{figure}

The results presented in Table \ref{table:Freq} seem to show that the proposed procedure is efficient to find the correct model from 150 sample size, but also it is acceptable with smaller sample size. When we comparing the percentages found for the mean and dispersion models, we found that in both the mean and dispersion models the scheme of variable selection is efficient, however for the dispersion model the efficiency is lower than that of the mean model, mainly for small samples. For the dispersion models, we had maximum percentage of Optimal model equal to $78.2 \%$, considering the $AIC_c$; while for the mean models, the maximum percentage was $98.9\%$, considering the criterion $\tilde{R}_m^{2}$ with $\lambda_n=\sqrt{n}$. 

When comparing the criteria $\tilde{R}_{m}^{2}$ ($\tilde{R}_{d}^{2}$) and $EAIC (AIC_c)$, used as selection measures, we note that, for the mean models, the criterion $\tilde{R}_{m}^{2}$ with $\lambda_n =\sqrt{n}$ was better and for the dispersion models, the $AIC_c$ performed better. These conclusions were observed for all sample sizes.

The graphs displayed in Figure \ref{fig:graphics-normal}, for the percentages shown in Table \ref{table:Freq}, show only the results for the Optimal and Type 1 models and visually confirm part of the conclusions obtained previously. The solid line show the percentage of hits of the proposed procedure using the criterion $\tilde{R}_{m}^2$ ($\tilde{R}_{d}^{2}$) with $\lambda_n=\log n$, the dashed line show the percentage of hits using the criterion $\tilde{R}_{m}^{2}$ ($\tilde{R}_{d}^{2}$) with $\lambda_n=\sqrt{n}$, the dotted line show the percentage of hits using the criterion $\tilde{R}_{m}^{2}$ ($\tilde{R}_{d}^{2}$) with $\lambda_n=1$ and the dash-dotted line presents the percentage of hits using the $EAIC (AIC_c)$. The graphs represent well the efficiency of the proposed procedure, since they show the growth of the percentages of the Optimal models and the considerable decrease of the percentages of the Type 1 models. With respect to the comparison of the criteria, the graphs referring to the mean model have the graph lines very close, with an advantage for the $\tilde{R}_{m}^{2}$ with $\lambda_n=\sqrt{n}$. In the graphs referring to the dispersion model, the $AIC_c$ is represented by the line farthest from the others, indicating a best efficiency than $\tilde{R}_{d}^2$, regardless of the $\lambda_n$ considered.

\subsection{Simulation study for Binomial distribution}
\label{SimBinomial}

For the mean model with Binomial probability distribution with index $m$ or Binomial type, i.e., with $V(\mu_i)=\mu_i(1-\mu_i/m)$, the data simulation process must ensure that $E(Y_i)=\mu_i$ and $Var(Y_i)=\phi_i \mu_i(1-\mu_i/m)$ and thus, assuring overdispersion. According to \cite{McCullaghNelder}, overdispersion occurs when the variance of the response exceeds the nominal variance of theoretical model. For the Binomial distribution the nominal variance is $\mu(1- \mu/m)$. Overdispersion can arise in a number of ways, but the simplest and most common mechanism is clustering in the population. The generation of data is adapted from \cite{McCullaghNelder}, p. 125. 

Clusters usually vary in size, but, in the same way as considered by \cite{McCullaghNelder}, we also assume that the cluster size, $k$, is fixed and that the $m$ individuals sampled come from $\ell=m/k$ clusters. Let's consider the total number of positive respondents, represented by the conditional random variable $Y|\Pi=S_1+\ldots+S_{\ell}$, where $S_j$, the number of positive respondents in the cluster $j$, with $j=1,\ldots,\ell$, are independent random variables with the same distribution as the random variable $S \sim Binomial(k, \pi)$. We also assume that the random variable $\Pi$ has a Beta distribution. Note that, for given $\pi_i$, $Y_i|\pi_i \sim Binomial(m,\pi_i)$. In this way,  for $\Pi_i \sim Beta(a_i,b_i)$, we can obtain $E(Y_i)= m\lambda_i$ and $Var(Y_i)=m\lambda_i(1-\lambda_i)[1+(m-1)\delta_i]$, where $\lambda_i=\frac{a_i}{a_i + b_i}$ and $\delta_i=\frac{1}{a_i + b_i +1}$. The random variable $Y_i$ has the Beta-Binomial distribution, see \cite{McCullaghNelder}, p. 140. Thus, by taking $\mu_i=m\lambda_i$ and $\phi_i=1+(m-1)\delta_i$ we have the desired result.

For the Binomial distribution, the values chosen for the parameter vector $(\beta_0,\beta_1,\beta_2,$ $\beta_3, \gamma_0, \gamma_1,\gamma_2,\gamma_3)$ used to perform simulations were $(0.2,0.6,0,0.8,$ $1,0,0,2.5)$, where the zero value means that the parameter was not used in the simulation. We have also considered $k=5$ and $m=10$. For each Monte Carlo simulation, the responses were generated considering log link function for the dispersion model and logistic link function for the mean model. 

\begin{table}[!h]
\scalefont{0.80}
\centering
\caption{Percentages of model types found in 1000 Monte Carlo simulations of the proposed variable selection procedure for the Binomial distribution}
\label{table:Freq3}
\begin{tabular}{ll|llllllcccccccccllllll}
\hline
                            &                            &  &  & \multicolumn{4}{c}{Optimal}                                                                         & \multicolumn{1}{l}{} & \multicolumn{1}{l}{} & \multicolumn{1}{l}{} & \multicolumn{4}{c}{Type 2}                                                                            & \multicolumn{1}{l}{} & \multicolumn{1}{l}{} &  & \multicolumn{4}{c}{Type 1}                                                                          &  \\ \cline{4-9} \cline{11-16} \cline{18-23} 
                            &                            &  &  & \multicolumn{4}{c}{Sample size}                                                                     & \multicolumn{1}{l}{} & \multicolumn{1}{l}{} & \multicolumn{1}{l}{} & \multicolumn{4}{c}{Sample size}                                                                       & \multicolumn{1}{l}{} & \multicolumn{1}{l}{} &  & \multicolumn{4}{c}{Sample size}                                                                     &  \\ \cline{5-8} \cline{12-15} \cline{19-22}
Model                       & Criterion                  &  &  & \multicolumn{1}{c}{25} & \multicolumn{1}{c}{50} & \multicolumn{1}{c}{100} & \multicolumn{1}{c}{150} &                      &                      &                      & 25                      & 50                      & 100                     & 150                     &                      &                      &  & \multicolumn{1}{c}{25} & \multicolumn{1}{c}{50} & \multicolumn{1}{c}{100} & \multicolumn{1}{c}{150} &  \\ \hline
\multirow{4}{*}{Mean}       & $\tilde{R}_{m}^2(\log n)$  &  &  & 33.9                   & 63.1                   & 66.1                    & 88.8                    &                      &                      &                      & 0.4                     & 2.1                     & 2.0                     & 1.4                     &                      &                      &  & 65.7                   & 34.8                   & 31.9                    & 9.8                     &  \\
                            & $\tilde{R}_{m}^2(\sqrt n)$ &  &  & 13.4                   & 51.8                   & 82.2                    & 92.4                    &                      &                      &                      & 0.4                     & 0.2                     & 0.3                     & 0.3                     &                      &                      &  & 86.2                   & 48.0                   & 17.5                    & 7.3                     &  \\
                            & $\tilde{R}_{m}^2(1)$       &  &  & 35.3                   & 69.0                   & 80.8                    & 87.7                    & \multicolumn{1}{l}{} & \multicolumn{1}{l}{} & \multicolumn{1}{l}{} & \multicolumn{1}{l}{2.9} & \multicolumn{1}{l}{3.4} & \multicolumn{1}{l}{4.7} & \multicolumn{1}{l}{4.6} & \multicolumn{1}{l}{} & \multicolumn{1}{l}{} &  & 61.8                   & 27.6                   & 14.5                    & 7.7                     &  \\
                            & $EAIC$                     &  &  & 53.9                   & 88.6                   & 94.4                    & 96.2                    &                      &                      &                      & 2.4                     & 4.3                     & 5.3                     & 0.0                     &                      &                      &  & 43.7                   & 7.1                    & 0.3                     & 3.8                     &  \\ \hline
\multirow{4}{*}{Dispersion} & $\tilde{R}_{d}^2(\log n)$  &  &  & 13.8                   & 34.6                   & 38.4                    & 80.8                    &                      &                      &                      & 0.2                     & 1.4                     & 1.3                     & 6.5                     &                      &                      &  & 86.0                   & 64.0                   & 60.3                    & 12.7                    &  \\
                            & $\tilde{R}_{d}^2(\sqrt n)$ &  &  & 11.9                   & 33.8                   & 66.2                    & 83.6                    &                      &                      &                      & 0.4                     & 1.8                     & 4.3                     & 4.9                     &                      &                      &  & 87.7                   & 64.4                   & 29.5                    & 11.5                    &  \\
                            & $\tilde{R}_{d}^2(1)$       &  &  & 11.2                   & 33.4                   & 65.5                    & 79.4                    & \multicolumn{1}{l}{} & \multicolumn{1}{l}{} & \multicolumn{1}{l}{} & \multicolumn{1}{l}{0.8} & \multicolumn{1}{l}{2.3} & \multicolumn{1}{l}{4.1} & \multicolumn{1}{l}{6.4} & \multicolumn{1}{l}{} & \multicolumn{1}{l}{} &  & 88                     & 64.3                   & 30.4                    & 14.5                    &  \\
                            & $AIC_c$                      &  &  & 17.3                   & 43.5                   & 69.7                    & 83.6                    & \multicolumn{1}{l}{} & \multicolumn{1}{l}{} & \multicolumn{1}{l}{} & \multicolumn{1}{l}{0.4} & \multicolumn{1}{l}{3.1} & \multicolumn{1}{l}{4.1} & \multicolumn{1}{l}{5.5} & \multicolumn{1}{l}{} & \multicolumn{1}{l}{} &  & 81,9                   & 53,4                   & 26,2                    & 10,9                    &  \\ \hline
\end{tabular}
\end{table}

\begin{figure}[!ht]
\centering
\includegraphics[scale=0.7]{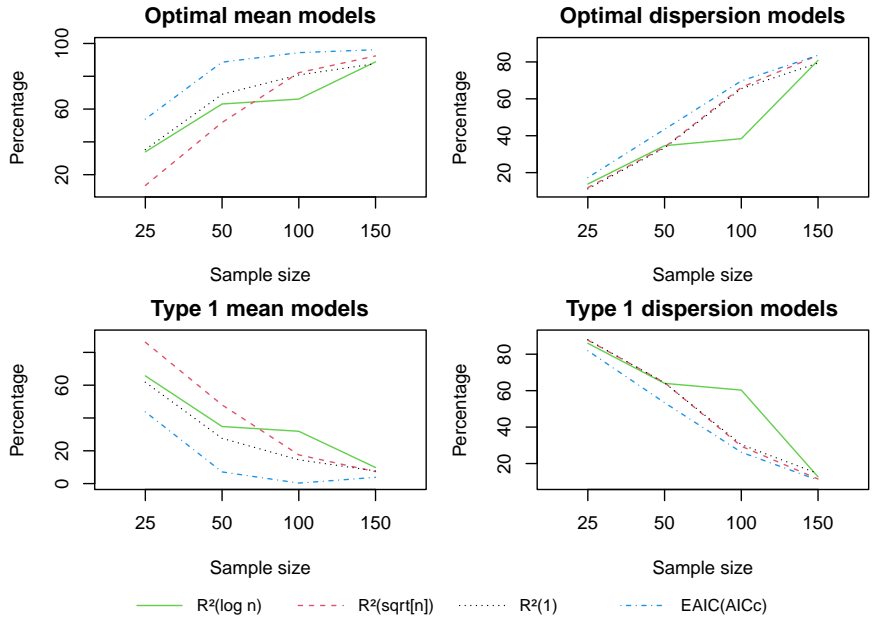}
\caption{Graphs, obtained from the simulations of the Binomial distribution, regarding the relationship between sample size and percentage found for the Optimal and Type 1 models, for the $\tilde{R}_{m}^{2}$, $\tilde{R}_{d}^{2}$, $EAIC$ and $AIC_c$ criteria, where the solid line represents the graph for $\tilde{R}_{m}^2$ $(\tilde{R}_{d}^2)$ with $\lambda_n=\log n$, the dashed line represents the graph for $\tilde{R}_{m}^2$ $(\tilde{R}_{d}^2)$, with $\lambda_n=\sqrt{n}$, the dotted line represents the graph for $\tilde{R}_{m}^2$ $(\tilde{R}_{d}^2)$, with $\lambda_n=1$ and the dash-dotted line represents the graph for $EAIC (AIC_c)$.}
\label{fig:graphics-binomial}
\end{figure}

In the Binomial case, the simulations had a low success rate for the Optimal models for 25 sample size, however there was a rapid growth of this rate, reaching excellent values for 150 sample size. This behavior was observed for both the mean and dispersion models.
Unlike the simulations for the Normal case, the $EAIC$ criterion showed better results for the mean model, however it is observed that the criteria converge to a similarity when the sample size is increased. For smaller samples, we observed a large difference between the criteria, indicating that in this case the $EAIC$ criterion must be chosen, but for large samples both criteria can be chosen. For the dispersion model, the $AIC_c$ criterion was slightly better than the other criteria, although in some cases the values were close. These results show that the choice of the criterion must be made with care.

The graphs displayed in Figure \ref{fig:graphics-binomial}, for the percentages shown in Table \ref{table:Freq3}, show what was discussed before, where we can see the lines, which are divergent for small samples, converging to a common point. The dash-dotted line is generally above the others in the Optimal models graphs and below the others in the Type 1 models graphs, indicating the best result for the $EAIC$ ($AIC_c$) criterion for small samples and the approximation of lines at 150 sample size indicating that the criteria converge to be similar.

\subsection{Simulation study for Poisson distribution}

For the mean model with Poisson probability distribution or Poisson type, i.e., with $V(\mu_i)=\mu_i$, the data simulation process must ensure that $E(Y_i)=\mu_i$ and $Var(Y_i)=\phi_i \mu_i$. The generation of data is adapted from \cite{McCullaghNelder}, p. 198. Consider the random variable $Y|N=S_1+\ldots+S_N$, where $S_j$ are independent and identically distributed random variables and $N$ is a random variable with Poisson distribution, independent of $S_j$, $\forall j=1,\ldots,N$. In this way we can obtain $E(Y_i)=E(N_i)E(S_{i})$    and  $Var(Y_i)=E(N_i)E(S_{i}^2)$. Note that there will be overdispersion if $Var(Y_i)>E(Y_i)$, that is, if $E(S_{i}^2)>E(S_{i})$. Thus, in order to obtain the desired result, it is sufficient to take $N_i \sim Poisson(\mu_i/ \rho_i)$ and  $S_i \sim Poisson(\rho_i)$ with $\rho_i=\phi_i-1 > 0$.

The values chosen for the parameter vector $(\beta_0,\beta_1,\beta_2,\beta_3,$ $\gamma_0,\gamma_1,\gamma_2,\gamma_3)$ used to perform simulations were $(1.5,3,2,0,0.2,0,3,0)$, where the zero value means that the parameter was not used in the simulation. For each Monte Carlo simulation, the responses were generated considering log link function for both mean and dispersion models.

\begin{table}[!h]
\scalefont{0.80}
\centering
\caption{Percentages of model types found in 1000 Monte Carlo simulations of the proposed variable selection procedure for the Poisson distribution}
\label{table:Freq2}
\begin{tabular}{ll|llllllcccllllccllllll}
\hline
                            &                            &  &  & \multicolumn{4}{c}{Optimal}                                                                         & \multicolumn{1}{l}{} & \multicolumn{1}{l}{} & \multicolumn{1}{l}{} & \multicolumn{4}{c}{Type 2}                                                                          & \multicolumn{1}{l}{} & \multicolumn{1}{l}{} &  & \multicolumn{4}{c}{Type 1}                                                                          &  \\ \cline{4-9} \cline{11-16} \cline{18-23} 
                            &                            &  &  & \multicolumn{4}{c}{Sample size}                                                                     & \multicolumn{1}{l}{} & \multicolumn{1}{l}{} & \multicolumn{1}{l}{} & \multicolumn{4}{c}{Sample size}                                                                     & \multicolumn{1}{l}{} & \multicolumn{1}{l}{} &  & \multicolumn{4}{c}{Sample size}                                                                     &  \\ \cline{5-8} \cline{12-15} \cline{19-22}
Model                       & Criterion                  &  &  & \multicolumn{1}{c}{25} & \multicolumn{1}{c}{50} & \multicolumn{1}{c}{100} & \multicolumn{1}{c}{150} &                      &                      &                      & \multicolumn{1}{c}{25} & \multicolumn{1}{c}{50} & \multicolumn{1}{c}{100} & \multicolumn{1}{c}{150} &                      &                      &  & \multicolumn{1}{c}{25} & \multicolumn{1}{c}{50} & \multicolumn{1}{c}{100} & \multicolumn{1}{c}{150} &  \\ \hline
\multirow{4}{*}{Mean}       & $\tilde{R}_{m}^2(\log n)$  &  &  & 88.3                   & 90.5                   & 91.1                    & 92.4                    &                      &                      &                      & 11.3                   & 9.3                    & 8.8                     & 7.5                     &                      &                      &  & 0.4                    & 0.2                    & 0.1                     & 0.1                     &  \\
                            & $\tilde{R}_{m}^2(\sqrt n)$ &  &  & 92.6                   & 92.6                   & 93.3                    & 95.3                    &                      &                      &                      & 7.00                   & 6.8                    & 6.5                     & 4.6                     &                      &                      &  & 0.4                    & 0.6                    & 0.2                     & 0.1                     &  \\
                            & $\tilde{R}_{m}^2(1)$       &  &  & 89.5                   & 90.7                   & 91.3                    & 90.8                    & \multicolumn{1}{l}{} & \multicolumn{1}{l}{} & \multicolumn{1}{l}{} & 10.4                   & 9.2                    & 8.7                     & 9.2                     & \multicolumn{1}{l}{} & \multicolumn{1}{l}{} &  & 0.1                    & 0.1                    & 0.1                     & 0.0                     &  \\
                            & $EAIC$                     &  &  & 85.9                   & 91.8                   & 91.3                    & 92.4                    &                      &                      &                      & 6.7                    & 6.6                    & 8.6                     & 7.5                     &                      &                      &  & 7.4                    & 1.6                    & 0.1                     & 0.1                     &  \\ \hline
\multirow{4}{*}{Dispersion} & $\tilde{R}_{d}^2(\log n)$  &  &  & 10.6                   & 27.3                   & 47.3                    & 58.3                    &                      &                      &                      & 0.4                    & 3.4                    & 9.5                     & 13.4                    &                      &                      &  & 89.0                   & 69.3                   & 43.2                    & 28.3                    &  \\
                            & $\tilde{R}_{d}^2(\sqrt n)$ &  &  & 13.5                   & 27.2                   & 51.6                    & 59.8                    &                      &                      &                      & 0.3                    & 2.3                    & 7.7                     & 13.1                    &                      &                      &  & 86.2                   & 70.5                   & 40.7                    & 27.1                    &  \\
                            & $\tilde{R}_{d}^2(1)$       &  &  & 11.2                   & 29.7                   & 46.0                    & 57.6                    & \multicolumn{1}{l}{} & \multicolumn{1}{l}{} & \multicolumn{1}{l}{} & 1.0                    & 3.2                    & 9.2                     & 15.6                    & \multicolumn{1}{l}{} & \multicolumn{1}{l}{} &  & 87.8                   & 67.1                   & 44.8                    & 28.8                    &  \\
                            & $AIC_c$                      &  &  & 51.7                   & 64.9                   & 78.8                    & 84.0                      &                      &                      &                      & 2.4                    & 3.1                    & 3.1                     & 4.7                     &                      &                      &  & 51,7                   & 32.0                   & 18.1                    & 11.3                    &  \\ \hline
\end{tabular}
\end{table}

\begin{figure}[!ht]
\centering
\includegraphics[scale=0.7]{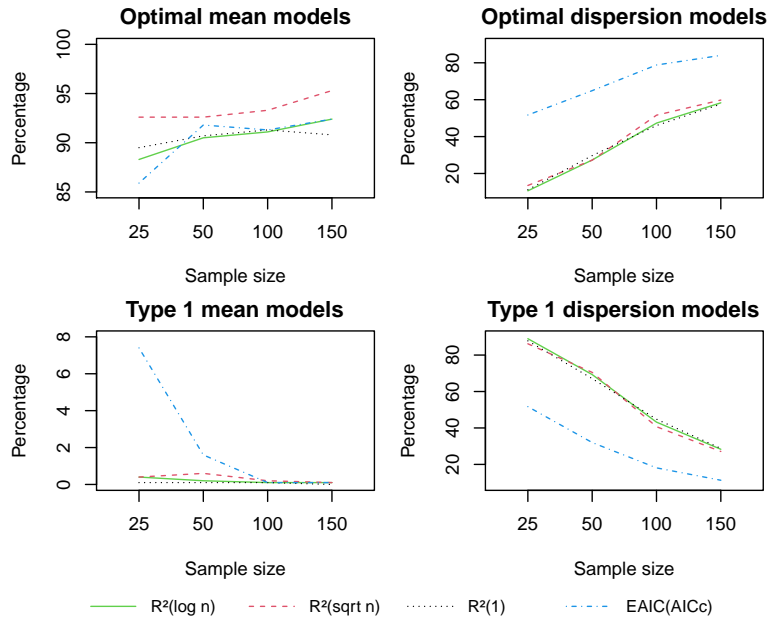}
\caption{Graphs, obtained from the simulations of the Poisson distribution, regarding the relationship between sample size and percentage found for the Optimal and Type 1 models, for the $\tilde{R}_{m}^{2}$, $\tilde{R}_{d}^{2}$ and $EAIC$ criteria,  where the solid line represents the graph for $\tilde{R}_{m}^2$ $(\tilde{R}_{d}^2)$ with $\lambda_n=\log n$, the dashed line represents the graph for $\tilde{R}_{m}^2$ $(\tilde{R}_{d}^2)$, with $\lambda_n=\sqrt{n}$, the dotted line represents the graph for $\tilde{R}_{m}^2$ $(\tilde{R}_{d}^2)$, with $\lambda_n=1$ and the dash-dotted line represents the graph for $EAIC (AIC_c)$.}
\label{fig:graphics-poisson}
\end{figure}

Table \ref{table:Freq2} shows that, for the mean models, the procedure for variable selection remains efficient for the Poisson distribution  and we even have higher efficiency for smaller samples compared to previous distributions. The $\tilde{R}_{m}^{2}$  with $\lambda_n=\sqrt{n} $ criterion had a slight advantage over the other criteria. 

For the dispersion models, the $AIC_c$ criterion was better, with an high rate of Optimal models for large samples. The $\tilde {R}_{d}^{2}$ criterion had an improvement in the Optimal models rate with the growth of the sample size, reaching close to $60 \%$. Also note that Type 1 models showed high rates for small samples, however there was an accelerated decrease in that rate when the sample size was increased.

In the graphs displayed in Figure \ref{fig:graphics-poisson}, we can see lines that are above the others, that is, for the mean models we observe the advantage of the $\tilde{R}_{m}^{2}$ criterion with $\lambda_n=\sqrt{n}$ (dashed line) and for the dispersion models we observe the advantage of the $AIC_c$ criterion (dash-dotted line).

For type 1 models, the $AIC_c$ criterion was better for dispersion models, with the dash-dotted line always below the others. For the mean models,
the criterion $\tilde{R}_{m}^{2}$, for any value of $\lambda_n$, was better than $EAIC$ for small samples, however the criteria become equivalent as the sample size increases.

\section{Application}
\label{Application}

The proposed variable selection procedure was applied to data from an injection molding experiment.

The experiment was performed to study the influence of seven controllable factors and three noise factors on the mean value and the variation in the percentage of shrinkage of products made by injection molding. Noise  factors  are fixed  during  the  experiment  but  are expected  to vary  randomly  outside  the  experimental  context.

The aim of the experiment was to determine the process parameter settings so that the shrinkage percentage was close to the target value and robust against environmental variations. The responses were percentages of shrinkage of products made by injection molding, the design for seven controllable factors ( $A=$ cycle  time,  $B=$ mould  temperature,  $C=$  cavity  thickness,  $D=$ holding  pressure,  $E=$ injection  speed,  $F=$  holding  time,  and  $G=$ gate  size) was a $2^{7-4}$ fractional factorial. At each setting of the  controllable factors, four observations were obtained from a $2^{3-1}$ fractional factorial with three noise factors ($M=$ percentage  regrind,  $N=$ moisture  content  and  $O=$ ambient temperature). The experimental data from the injection molding are shown in Table \ref{Tab_data_example}. 

The data set considered is well known in the literature of industrial experiments and has been analyzed by several authors such as \cite{Engel}, \cite{EngelHuele} and \cite{LeeNelderI}. Although the articles by \cite{EngelHuele} and \cite{LeeNelderI} have used the JMMD, the variable selection process was not discussed. In our approach, in addition to obtaining estimates of the parameters of the mean and dispersion models by using JMMD, we also built a consistent investigative procedure to verify the adequacy of the final joint model found. For the application of Algorithm 2, we considered $V(\mu)=1$ and identity link function for the mean model. For the dispersion model we considered the Gamma distribution with logarithmic link function. The joint models we considered for mean and dispersion were the same as those proposed by \cite{EngelHuele} and \cite{LeeNelderI}.

The selection measure considered in Algorithm 1 for the mean model was $\tilde R_m^2$ with $\lambda_n=\sqrt{n}$  and for the dispersion model the selection measure was the $AIC_c$. As explained in Section \ref{NumEvaluation},  the level of significance for the hypothesis tests performed was $0.10$.

Table \ref{table_Steps_Algorithm} presents the results of all steps of the variable selection process (Algorithm 2) taken to obtain the final joint model for the mean and dispersion. In each iteration of Algorithm 2, Algorithm 1 was used to find the terms of the mean and dispersion models. The additional terms to the current models, which appear in sequence in Table 6, are those that, for all terms not belonging to the current model, have the highest $\tilde R_m^2$ value, in case of the mean model, and the smallest $AIC_c$, in the case of the dispersion model. For these terms, which were candidates to enter the current model, Chi-square hypothesis tests, in the case of the dispersion model, and $F$ test, in the case of the mean model, were used to verify if, in fact, the term chosen must be added to the current model. Measurements $D$ and $D^{*}$ were used to obtain test statistics. The values of
$D$, $D^{*}$, $\chi^2$, $F$ and $\tilde R_m^2$, with $\lambda_n=\sqrt{n}$ and with $\lambda_n=1$, are shown for each step of the Algorithm 1, where $D$ is the deviance for the Gamma model,  $D^{*}=\sum_{i=1}^{n}\frac{1}{\phi_i}d_{i}^{*}$ (see Algorithm 1), $\chi^2$, the test statistic for the Chi-square test in the dispersion model, is the deviance difference for nested Gamma models and $F$ is the test statistic, given by equation (\ref{eqn:F}), for the $F$ test in the mean model. It is worth mentioning that the measure $\tilde R_m^2$ with $\lambda_n=1$, which was not used as a selection measure, was presented in Table \ref{table_Steps_Algorithm} only to show the impact on the value of $\tilde R_m^2$, when the number of terms in the model increases, considering the penalty with $\lambda_n=\sqrt{n}$ and with $\lambda_n=1$. In each iteration, the final models for both mean and dispersion are shown in bold.

Following Algorithm 2, since the value of $\tilde R_m^2$ for the mean model, obtained in the third iteration, was lower than that obtained in the second iteration, the procedure stopped. In this way, three iterations of Algorithm 2 were performed, obtaining at the end, in the second iteration, a model for the mean with $\tilde R_m^2(\sqrt{n})$ equal to 0.803 and $\tilde R_m^2(1)$ equal to 0.973. 
All the mean models found in the three iterations of Algorithm 2 had the same terms (see Table \ref{table_Steps_Algorithm}), differing only in the value of the effect estimates. Table \ref{Tab_Examp_Mean_Disp} presents the final estimates, with their standard deviations and Wald tests. Note that all estimates were significant at the 1\% level, except the terms $E$, $N$ and $C$ in the mean model, which entered in the model only after the variable selection procedure ends, due to the $CN$ and $EN$ interaction terms being significant.

\begin{figure}[!ht]
\centering
\includegraphics[scale=0.7]{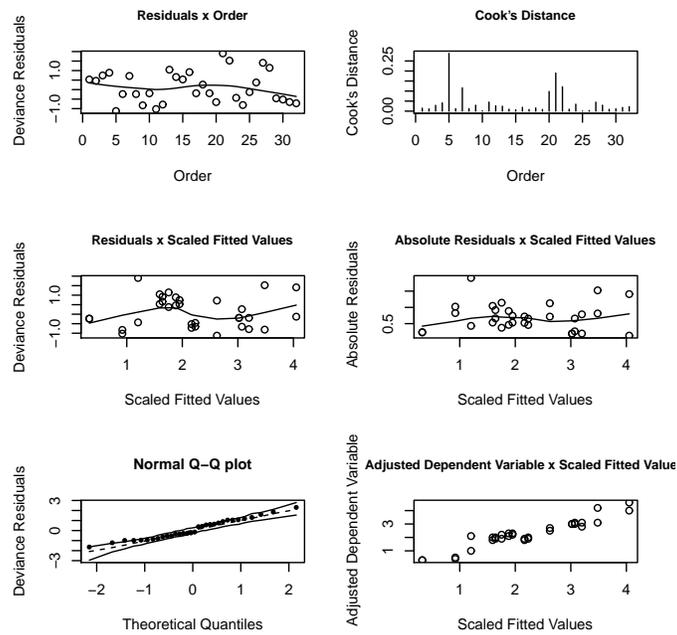}
\caption{ Diagnostic plots for mean model obtained from injection molding data. }
\label{fig:graphics-resid-mean}
\end{figure}

The diagnostic graphs for the mean model, displayed in Figure \ref{fig:graphics-resid-mean}, show that the model obtained is adequate, since it showed no tendency to violate the assumptions of independence, constant variance and normality of the residuals, there is no evidence of inadequate of the link and variance functions or points of influence.

\begin{figure}[!ht]
\centering
\includegraphics[scale=0.5]{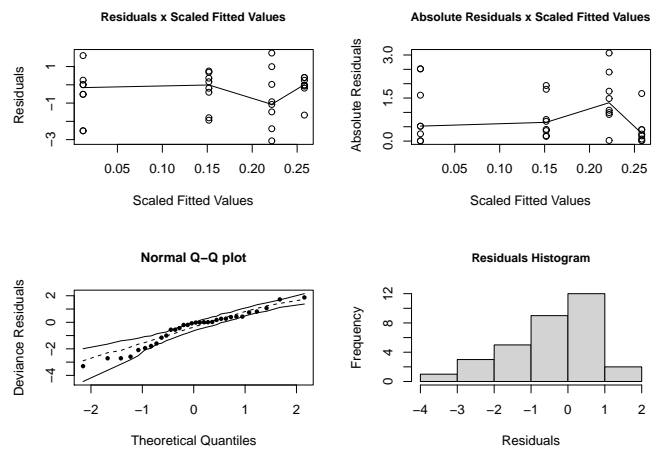}
\caption{Diagnostic plots for dispersion model obtained from injection molding data. }
\label{fig:graphics-resid-disp}
\end{figure}

For the dispersion model, the models obtained in each iteration of Algorithm 2 were different. This is due to the fact that the response variable, $d^*$, is not the same in each iteration of Algorithm 2. The final dispersion model had a value of $AIC_c$ equal to $-74.2471$ (see Table \ref{table_Steps_Algorithm}) and showed highly significant effects by the Wald test (see Table \ref{Tab_Examp_Mean_Disp}). Diagnostic graphs for the dispersion model are presented in Figure \ref{fig:graphics-resid-disp}.

\begin {table}[htb]
\caption{Experimental data from the injection molding}
\label{Tab_data_example}
\scriptsize
\centering
\begin{tabular}{llrrrrrrrrrrrrr}
\hline                          
& & & & & & & & & & \multicolumn{5}{c}{\scriptsize{Noise factors}} \\
\cline{11-15}
&\multicolumn{9}{c}{} & $M$ & $-1$ & $-1$ &  1   & 1 \\
&\multicolumn{9}{c}{} & $N$ & $-1$ &  1   & $-1$ & $1$ \\
&\multicolumn{8}{c}{\scriptsize{Controllable factors}} & & $O$ & $-1$ &  1 &  1 & $-1$ \\
\cline{3-9}
$\textrm{run}$ & & $A$ & $B$ & $C$ & $D$ & $E$ & $F$ & $G$ &  &  &  &  &  &  \\
\hline
1 & &$-1$ &$-1$ &$-1$ &$-1$ &$-1$ &$-1$ &$-1$ &  & & 2.2 & 2.1 & 2.3 & 2.3\\  
2 & &$-1$ &$-1$ &$-1$ & 1   & 1   & 1   & 1   &  & & 2.5 & 0.3 & 2.7 & 0.3\\  
3 & &$-1$ & 1   & 1   &$-1$ &$-1$ & 1   & 1   &  & & 0.5 & 3.1 & 0.4 & 2.8\\  
4 & &$-1$ & 1   & 1   & 1   & 1   &$-1$ &$-1$ &  & & 2.0 & 1.9 & 1.8 & 2.0\\  
5 & & 1   &$-1$ & 1   &$-1$ & 1   &$-1$ & 1   &  & & 3.0 & 3.1 & 3.0 & 3.0\\  
6 & & 1   &$-1$ & 1   & 1   &$-1$ & 1   &$-1$ &  & & 2.1 & 4.2 & 1.0 & 3.1\\  
7 & & 1   & 1   &$-1$ &$-1$ & 1   & 1   &$-1$ &  & & 4.0 & 1.9 & 4.6 & 2.2\\  
8 & & 1   & 1   &$-1$ & 1   &$-1$ &$-1$ & 1   &  & & 2.0 & 1.9 & 1.9 & 1.8\\  
\hline
\end{tabular}
\end{table}

\begin{table}[htb]
\scalefont{0.80}
\centering
\caption{Steps of variable selection algorithm for JMMD applied  to  data  from  the injection  molding  experiment}
\label{table_Steps_Algorithm}
\setlength\extrarowheight{-2.4pt} 
\begin{tabular}{llccccc} 
\hline
 \hline
\multicolumn{1}{l|}{Iteration}         & Dispersion model & $AIC_c$                        & $D$ $^\ddag$          & $\chi^2_c$ value$^\star$ & $Pr(>\chi^2_c)$ &            \\ \hline
\multicolumn{1}{l|}{\multirow{8}{*}{1}}  & 1                &  -                            &  -                   &  -                        &  -               &            \\ \cline{2-7} 
\multicolumn{1}{l|}{}                    & Mean model       & $\tilde{R}_{m}^2 (\sqrt{n})$ & $\tilde{R}_{m}^2 (1)$ & $D^{*}$ $^\ddag$         & $F_c$ value     & $Pr(>F_c)$ \\ \cline{2-7} 
\multicolumn{1}{l|}{}                    & 1                & -                       & -                & 34.6839                  & -          & -     \\
\multicolumn{1}{l|}{}                    & 1+CN             & -0.0060                      & 0.3063                & 24.0587                  & 13.2491         & 0.0010     \\
\multicolumn{1}{l|}{}                    & 1+CN+EN          & 0.2232                       & 0.5974                & 13.9628                  & 20.9687         & 0.0001     \\
\multicolumn{1}{l|}{}                    & 1+CN+EN+A        & 0.3234                       & 0.7735                & 7.8557                   & 21.7672         & 0.0001     \\
\multicolumn{1}{l|}{}                    & \bf{1+CN+EN+A+D}      & -0.0782                      & 0.8516                & 5.1467                   & 14.2120         & 0.0008     \\ \hline \hline
\multicolumn{1}{l|}{Iteration}         & Dispersion model & $AIC_c$                        & $D$ $^\ddag$          & $\chi^2_c$ value$^\star$ & $Pr(>\chi^2_c)$ &            \\ \hline
\multicolumn{1}{l|}{\multirow{12}{*}{2}} & 1                & -                       & 80.8899               & -                   & -          &            \\
\multicolumn{1}{l|}{}                    & 1+E              & -64.1778                     & 72.0353               & 4.4273                   & 0.0354          &            \\
\multicolumn{1}{l|}{}                    & 1+E+B            & -68.4785                     & 61.7544               & 5.1404                   & 0.0234          &            \\
\multicolumn{1}{l|}{}                    & \bf{1+E+B+G}          & -74.2471                     & 50.8136               & 5.4704                   & 0.0193          &            \\
\multicolumn{1}{l|}{}                    & 1+E+B+G+D        & -77.5280                     & 44.3609               & 3.2263                   & 0.0725          &            \\ \cline{2-7} 
\multicolumn{1}{l|}{}                    & Mean model       & $\tilde{R}_{m}^2 (\sqrt{n})$ & $\tilde{R}_{m}^2 (1)$ & $D^{*}$ $^\ddag$         & $F_c$ value     & $Pr(>F_c)$ \\ \cline{2-7} 
\multicolumn{1}{l|}{}                    & 1                & -                       & -               & 1,098.102                & -           & -    \\
\multicolumn{1}{l|}{}                    & 1+A              & 0.133                        & 0.402                 & 699.342                  & 17.106          & 0.0003     \\
\multicolumn{1}{l|}{}                    & 1+A+CN           & 0.378                        & 0.678                 & 404.023                  & 21.197          & 0.0001     \\
\multicolumn{1}{l|}{}                    & 1+A+CN+EN        & 0.841                        & 0.947                 & 52.454                   & 187.666         & 0.0000     \\
\multicolumn{1}{l|}{}                    & \bf{1+A+CN+EN+D}      & 0.803                        & 0.973                 & 26.624                   & 26.195          & 0.0000     \\  \hline \hline
\multicolumn{1}{l|} {Iteration}                              & Dispersion model & $AIC_c$                        & $D$ $^\ddag$          & $\chi^2_c$ value$^\star$ & $Pr(>\chi^2_c)$ &            \\ \hline
\multicolumn{1}{l|}{\multirow{10}{*}{3}} & 1                & -                       & 75.4049               & -                  & -          &            \\
\multicolumn{1}{l|}{}                    & \bf{1+D}              & 42.5318                      & 65.4232               & 4.9908                   & 0.0255          &            \\
\multicolumn{1}{l|}{}                    & 1+D+F            & 41.9160                      & 61.3397               & 2.0417                   & 0.1530          &            \\ \cline{2-7} 
\multicolumn{1}{l|}{}                    & Mean model       & $\tilde{R}_{m}^2 (\sqrt{n})$ & $\tilde{R}_{m}^2 (1)$ & $D^{*}$ $^\ddag$         & $F_c$ value     & $Pr(>F_c)$ \\ \cline{2-7} 
\multicolumn{1}{l|}{}                    & 1                & -                       & -                & 59.6112                  & -          & -     \\
\multicolumn{1}{l|}{}                    & 1+CN             & -0.0078                      & 0.3051                & 41.8419                  & 12.7404         & 0.0012     \\
\multicolumn{1}{l|}{}                    & 1+CN+EN          & 0.2266                       & 0.5992                & 24.3629                  & 20.8059         & 0.0001     \\
\multicolumn{1}{l|}{}                    & 1+CN+EN+A        & 0.5173                       & 0.8384                & 9.4400                   & 44.2627         & 0.0000     \\
\multicolumn{1}{l|}{}                    & \bf{1+CN+EN+A+D}      & 0.2260                       & 0.8935                & 6.2942                   & 13.4942         & 0.0010     \\ \hline
\multicolumn{7}{l}{$^{\ddag}$\tiny{$D$ is the deviance for the Gamma model and $D^{*}=\sum_{i=1}^{n}\frac{1}{\phi_i}d_{i}^{*}$.}}\\
\multicolumn{7}{l}{$^{\star}$\tiny{$\chi^2_c$ is the value of the Chi-square statistic, calculated from the difference in deviations }} \\ 
\multicolumn{7}{l}{\tiny{for two nested Gamma models.}} \\
\multicolumn{7}{l}{$^{\diamond}$\tiny{$F_c$ is the value of the F-test statistic, calculated by the equation (\ref{eqn:F}).}}

\end{tabular}
\end{table}

\def\tablename{Table}
\begin{table} [!htp]
\scalefont{0.85}
\begin{center}
\caption{Regression coefficients and  Wald test for the mean and dispersion models obtained from injection molding data} \vspace{0.3cm}
\begin{tabular}{lrrrrr} \hline
Mean model       &              &             &            &                 \\ \hline \hline
Terms            &  Estimate    &  Std. Error & $t$ value  & $Pr(>|t|)$      \\ \hline 
  1      &  2.24903 &   0.03322  & 67.693 & 0.0000 \\
 $A$      &  0.42802 &   0.06575  & 6.509 & 0.0000 \\
 $C$       &  0.07172 &   0.05802  & 1.236 & 0.2279   \\ 
 $N$      & -0.00433 &   0.05994 & -0.072 & 0.9430   \\
 $D$      & -0.28639 &   0.06407 & -4.470 & 0.0001 \\
 $E$      &  0.06528 &   0.05971  &  1.093 & 0.2851 \\
 $CN$      & 0.58684  &  0.03322  & 17.663 & 0.0000 \\
 $EN$      & -0.55727 &   0.05994 & -9.297 & 0.0000  \\\hline
Dispersion model &              &             &            &                 \\ \hline \hline
Terms            &  Estimate    &  Std. Error & $t$ value  & $Pr(>|t|)$      \\ \hline
1           &   -2.2973  &   0.1754  & -13.097 & 0.0000 \\
$E$         &   -0.8670  &   0.1754  & -4.942  & 0.0000 \\
$B$         &    0.6773  &   0.1754  &  3.861  & 0.0006 \\
$G$         &   -0.6015  &   0.1754  & -3.429  & 0.0019 \\ \hline 
\end{tabular}
\label{Tab_Examp_Mean_Disp}
\end{center}
\end{table}

\section{Conclusion}
\label{Conclusion}

The article presents a procedure for variable selection in joint modeling of the mean and dispersion with application in robust design experiments. The proposed procedure provides an iterative mechanism for selecting variables in the JMMD. The process takes advantage of the seesaw structure from the algorithm to estimate the parameters in the JMMD, where the weights from the dispersion model are incorporated into the mean model. For both mean and dispersion models, a selection measure is used to choose, in each iteration of the procedure, the variables that contribute to the best fit of the model. Appropriate hypothesis tests are used to confirm the choice of these variables or to terminate the procedure. As the weights from the dispersion model are incorporated into the mean model, the adjustment of the mean model becomes better. The final joint model obtained is the one that has the best fit for the mean model. In this way, the dispersion model is constituted by the terms that provide the best weights for the mean model.

For each of the mean and dispersion models, two alternative selection measures were used in the simulation studies. The $\tilde{R}_{m}^{2}$ and $EAIC$ selection measures were used in the mean model and $\tilde{R}_{d}^{2}$ and $AIC_c$ were used in the dispersion model. Each of the $\tilde{R}_{m}^{2}$ and $\tilde{R}_{d}^{2}$ criteria were used considering the variations based on the choices of $\lambda_n$ equal to $\sqrt{n}$, $\log n$ and 1.

In the simulation studies, performed for the Normal, Poisson and Binomial distributions, the $\tilde{R}_{m}^{2}$ criterion with $\lambda_n=\sqrt{n}$ was better for the Normal and Poisson cases, while the $EAIC$ was better for the Binomial case. The $AIC_c$ was better in all cases considered.

The efficiency of the proposed procedure depends on the criterion used as a selection measure and the sample size. As should be expected, the efficiency grows with the increase of the sample size. The simulations showed that, for a large sample size, the proposed procedure considerably reduces the risk of finding poor models and ensures that satisfactory models are obtained for mean and dispersion. However, the procedure proved to be reasonably good for small samples.

In application, discussed in Section \ref{Application}, the procedure for selecting variables was explained step by step and the results obtained for the mean and dispersion models were satisfactory, indicating the effectiveness of the proposed methodology and corroborating the results found by the simulations.

\begin{acks}[Acknowledgments]
The authors would like to thank the anonymous referees and the Editor for their constructive comments that improved the quality of this paper.
\end{acks}

\begin{funding}
This work was supported by Research Support Foundation of the State of Minas Gerais (FAPEMIG) under Grant Number APQ-03365-18.

\end{funding}

{}

\end{document}